\def\tipo{2}

\ifnum \tipo = 2
 \documentstyle[twocolumn,aps,psfig]{revtex}
 
 \def\figsiz1{8.5cm}
 \def \frontmatter{\twocolumn[\hsize\textwidth\columnwidth\hsize\csname@twocolumnfalse\endcsname}

\else
 \documentstyle[preprint,aps,prl,psfig]{revtex}
 
 \def\figsiz1{12cm}
 \def\frontmatter{}
 
\fi

\begin{document}


\draft

\def\intl#1#2{\int\limits_{#1}^{#2}}

\frontmatter

\title{\bf Nonlinear Volatility of River Flux Fluctuations}

\date{\today}

\author{%
Valerie N. Livina$^1$, 
Yosef Ashkenazy$^2$,
Peter Braun$^3$,
Roberto Monetti$^4$, 
Armin Bunde$^5$,
Shlomo Havlin$^1$
}

\address{$^1$Minerva Center and Department of Physics, 
Bar-Ilan University, Ramat-Gan 52900, Israel \\
$^2$Dep. of Earth, Atmospheric and Planetary Sciences,
Massachusetts Institute of Technology, Cambridge, MA 02139, USA\\
$^3$Bayerisches Landesamt f\"ur Wasserwirtschaft,
Lazarettstr. 67, D-80636 
M\"unchen, Germany \\
$^4$Center for Interdisciplinary Plasma Science (CIPS),
Max-Planck-Institut f\"ur Extraterrestrische Physik,
Giessenbachstr.~1, 85749 Garching, Germany 
\\
$^5$Institut f\"ur Theoretische Physik III, 
Justus-Liebig-Universit\"at Giessen,
Heinrich-Buff-Ring 16, 35392 Giessen, Germany\\
}

\date{\today}
\maketitle

\begin{abstract}
We study the spectral properties of the magnitudes of river flux
increments, the volatility. 
The volatility series exhibits (i) strong seasonal periodicity and
(ii) strongly power-law correlations for time scales less than one year. 
We test the nonlinear properties of the river flux increment series by
randomizing its Fourier phases and find that the surrogate volatility
series (i) has almost no seasonal periodicity and (ii) is weakly
correlated for time scales less than one year. 
We quantify the degree of nonlinearity by measuring (i) the amplitude
of the power spectrum at the seasonal peak and (ii) the
correlation power-law exponent of the volatility series.
\end{abstract} 

\ifnum \tipo = 2
]
\fi

\def\figureI{
\begin{figure}[thb]
\centerline{\psfig{figure=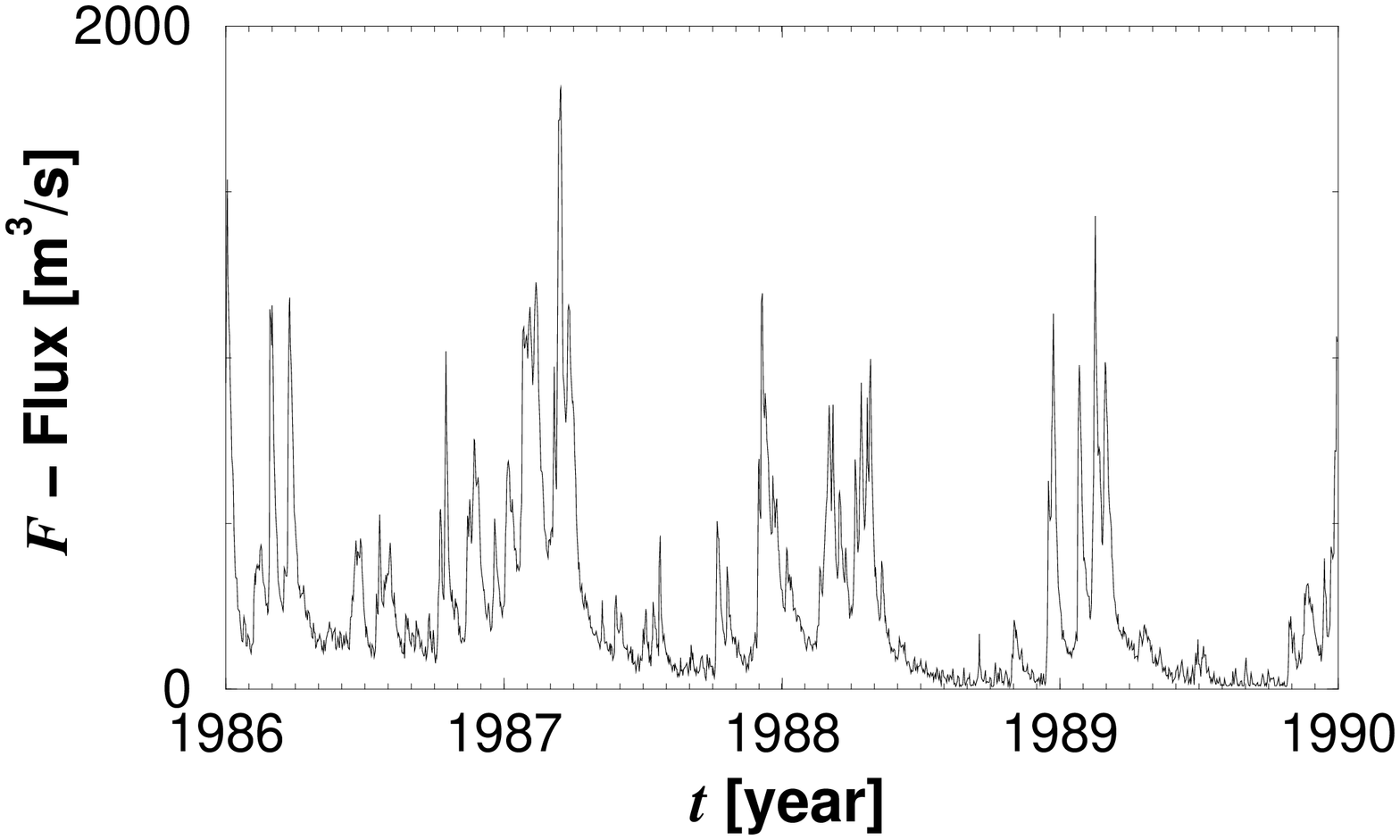,width=\figsiz1,angle=-90}}
{
\ifnum\tipo=2
\vspace*{0.0truecm}
\fi
\caption{\label{fig1} 
Typical river flow time series of Maas river (Europe, 1986-1990). The
record shows a periodic pattern with irregular fluctuations. Note the
large fluctuations around large river flow compare to small
fluctuations around small river flow. 
}}
\end{figure}
}

\def\figureII{
\begin{figure}[thb]
\centerline{\psfig{figure=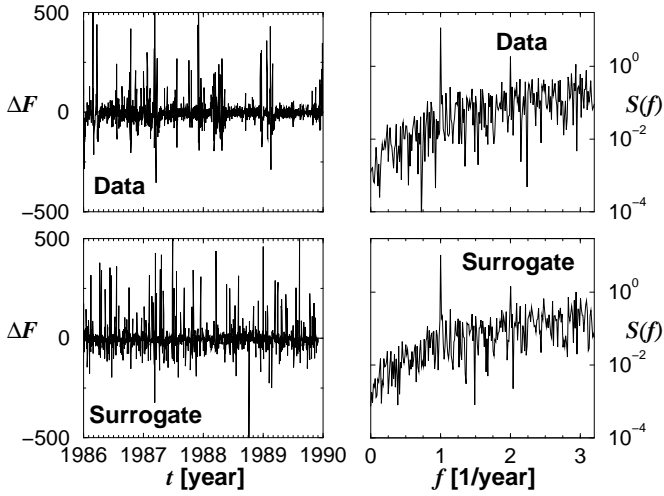,width=\figsiz1,angle=-90}}
{
\ifnum\tipo=2
\vspace*{0.0truecm}
\fi
\caption{\label{fig2} 
River flux increment series of Maas river (left panels) and their
corresponding power spectra (right panels) before (upper panels) and
after (lower panels) the surrogate test for nonlinearity. The
series length is 80 years where just the last 4 years of the record
are shown in the left panels. The original river flow increment series
and the surrogate increment series have identical probability
distribution and very similar power spectrum.  
}}
\end{figure}
}

\def\figureIII{
\begin{figure}[thb]
\centerline{\psfig{figure=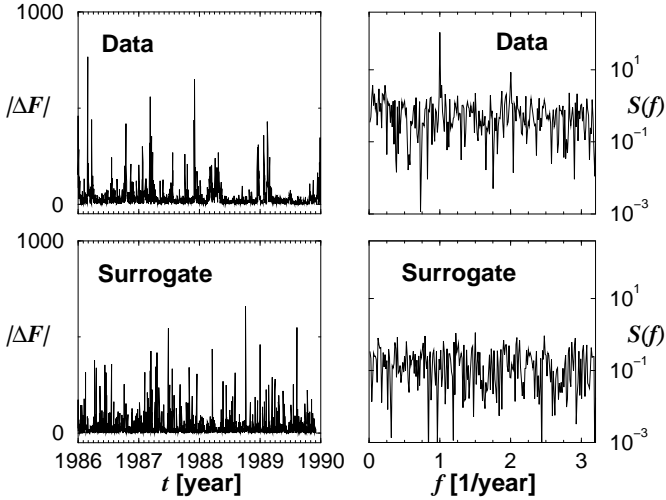,width=\figsiz1,angle=-90}}
{
\ifnum\tipo=2
\vspace*{0.0truecm}
\fi
\caption{\label{fig3} 
Same as Fig. \protect\ref{fig2} but for the river flow volatility
series, $|\Delta F_i|$. Here, the original volatility series shows a
pronounced seasonal peak while the surrogate volatility series doesn't
show such a peak indicating that the periodicity in the volatility
series is a result of nonlinearity. 
}}
\end{figure}
}

\def\figureIV{
\begin{figure}[thb]
\centerline{\psfig{figure=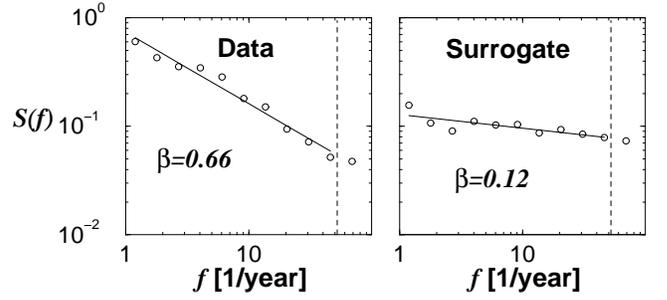,width=\figsiz1,angle=-90}}
{
\ifnum\tipo=2
\vspace*{0.0truecm}
\fi
\caption{\label{fig4} 
Log-log plot of the power spectra shown in
Fig. \protect\ref{fig3}. The solid lines are the best fit of $S(f)
\sim 1/f^\beta$ for frequencies $1.05yr^{-1} < f < 52yr^{-1}$; we use
logarithmic binning for the exponent calculation. The
original volatility series (left panel) decays as a power law
$1/f^{\beta=0.66}$ indicating long-range correlations of the volatility
series. The power spectrum of the linearized surrogate volatility
series has a flater spectrum indicating much less correlated
behavior. Thus, correlations in the volatility are an additional
measure for nonlinearity of the river flow increment time series.  
The vertical dashed lines indicate 1 week periodicity.
}}
\end{figure}
}

\def\figureV{
\begin{figure}[thb]
\centerline{\psfig{figure=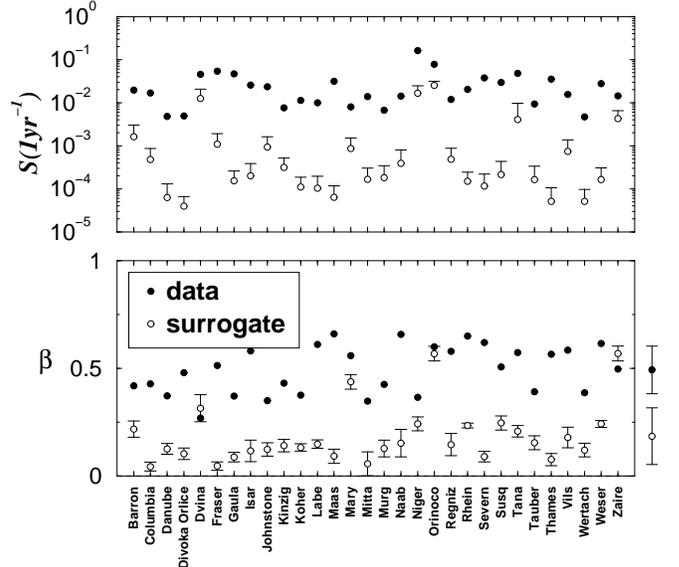,width=\figsiz1,angle=-90}}
{
\ifnum\tipo=2
\vspace*{0.0truecm}
\fi
\caption{\label{fig5} 
A summary of the results obtained for the 30 world rivers. For each
river flow increment series ($\bullet$) we generated 10 surrogate
series ($\circ$) and calculated the amplitude of the
seasonal peak of the volatility series (upper panel) and the scaling
exponent $\beta$ for frequencies $1.05yr^{-1} < f < 52yr^{-1}$ (lower
panel); 
the average and 1 standard deviation are shown. In order to
systematically compare the results of the different rivers we subtract
from the volatility series its mean and normalized by its standard
deviation. The seasonal peak of the volatility series is significantly
higher compare to the seasonal frequency of the surrogate volatility
series (upper panel). The scaling exponent $\beta$ shown in the lower
panel is systematically higher for the original volatility series. 
For 28 rivers the original volatility exponent is larger than its surrogate 
exponent where 27 of these 28 exponents lie well above the error
bars. The error bars on the right hand side are the group average
$\pm$ 1 standard deviation for the original and surrogate volatility
scaling exponent.
}}
\end{figure}
}

\def\figureVII{
\begin{figure}[thb]
\centerline{\psfig{figure=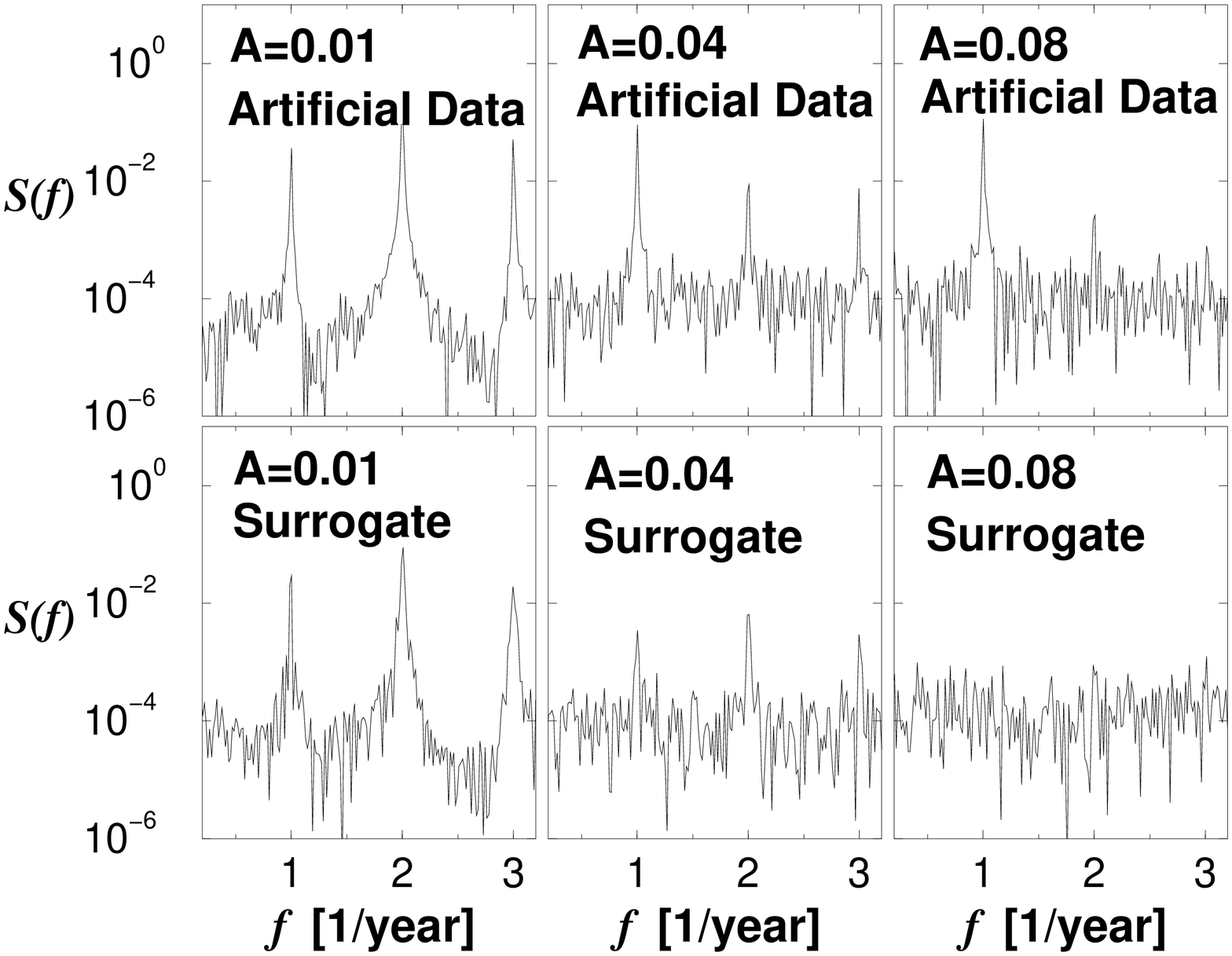,width=\figsiz1,angle=-90}}
{
\ifnum\tipo=2
\vspace*{0.0truecm}
\fi
\caption{\label{fig7} 
The power spectrum of the normalized volatility series, $|\Delta x_i|$,
of an artificial time series $x_i=(1+A\eta_i)s_i$ for different noise
level $A$; see Eqs. (\protect\ref{e1}),(\protect\ref{e2}). The upper
panels shows the power spectra of the original 
volatility series and the lower panels show the power spectra of the
surrogate volatility series. When the noise level increases (from the left
panels to the right panels) the seasonal peak of the surrogate
volatility series reduces. The harmonics of the power spectra are
partly caused by 
the asymmetric $x_i$ and partly because of the absolute value
operation for the volatility series.
}}
\end{figure}
}



Climate is strongly forced by the periodic variations of Earth with
respect to state of the solar system. The seasonal variations in the
solar radiation cause to periodic changes in temperature and
precipitation which eventually lead to seasonal periodicity of river
flow. In spite of this well defined seasonal change, river flow
exhibits highly unpredictable complex behavior; floods and droughts
are usually unexpected and cause severe damage in life,
housing, and agriculture products. Hence, river flow is likely to have
indirect nonlinear response to the seasonal changes in solar
radiation. 

Many components of  the water budget of a catchment
are coupled in a nonlinear fashion. The key for all interactions between
atmospheric processes like precipitation, temperature, humidity (or
extraterrestrical inputs like sun radiance) and  surface runoff is
the soil. The dynamic state of this key variable is highly
nonlinear. 

By means of the proposed methods in this paper it will be possible to
characterize quantitativly the degree of nonlinearity of the involved
processes by investigating the outputs of the catchment
(the resulting flux time series) only. This 
nonlinearity test would be very
helpful both for the design of time series models and
statistical prediction algorithms.

There are several statistical features found in the 
earlier studies of river flow
fluctuations. E.g., river flow fluctuations have broad probability
distribution \cite{murdock,vogel}. Moreover, river flow fluctuations
have unique temporal organization; they are long-range power-law
correlated and possess scale invariant structure 
\cite{hurst,turcotte}.
These river flow power-law correlations are usually characterized by
scaling exponent \cite{tessier,pandey} as was originally defined by
Hurst in his seminal work \cite{hurst} regarding the Nile river
floodings. However, such scaling laws only quantify the linear
properties (two-point correlations) of a time series. Here we
study other nonlinear statistical aspects of river flow fluctuations.


A nonlinearity of a stationary time series may be defined with respect
to its Fourier phases \cite{Schrieber00,ash_PRL00}. Series that its
statistical properties are independent of the Fourier phases may be
defined as {\it linear} otherwise the series may be defined as {\it
nonlinear}. Autoregression processes and fractional Brownian motion
are examples for linear processes while multifractal processes 
are examples for nonlinear
processes. Recently it has been shown that volatility correlations of
long-range power-law correlated time series reflects the degree of
nonlinearity of a time series \cite{ash_PRL00}. Given a time series,
$x_i$, the 
volatility series is defined as the magnitudes of the series
increments, $|\Delta x_i| \equiv |x_{i+1}-x_i|$. It was found that
long-range correlated linear series has uncorrelated volatility series
while long-range correlated nonlinear series has correlated volatility
series; see \cite{ash_PRL00} for details. Power-law correlations in the
volatility series indicate that the magnitudes $|\Delta x_i|$ are
clustered into self-similar patches of small and big magnitudes --- a
big magnitude increment is likely to precede a big magnitude increment
and vice versa. When the volatility series, $|\Delta x_i|$, is
uncorrelated the increment series is homogeneous. Volatility
correlations were found, for example, in econometric time series
\cite{Liu99}, heart interbeat interval series \cite{ash_PRL00,jan02},
and human gait dynamics \cite{GaitAshkenazy}.  

Here we study the volatility properties of river flow
fluctuations. We first extend the notion of volatility to periodic
time series by measuring the periodicity of the volatility series. We
find that after randomizing the Fourier phases of the 
river flow increment series, the periodicity of the volatility series
is almost diminished indicating that ``periodic volatility'' is a
result of nonlinearity. We also find long-range volatility correlations for
time scales below one year. Our results suggest that clusters of
magnitudes of river flow increments appear in two ways: periodic
clustering and long-range self-similar clustering. 

We analyze the daily river flux time series of 30 world rivers
scattered around the globe. The mean flux of these rivers
ranges from $\sim 0.6 m^3/s$ to $\sim 2 \times 10^5 m^3/s$ and thus
covers more than 5 orders of magnitudes. The length of the these time
series ranges 
from 26 years to 171 years with average length of 81.3 years. In
Fig. \ref{fig1} we present a typical example of 4 years (1986-1990) of river
flow data of the Maas river in Europe. It is evident that fluctuations
around large river flow are also large while the fluctuations around
small river flow are small.

To study the nonlinear properties of the river flow time series we apply
a surrogate data test to the river flow increment series. We use a
surrogate data test that preserves both the power spectrum and the
probability distribution of the river flow increment series
\cite{Schrieber00}. On the other hand, the Fourier phases of the
surrogate series are random. Thus, the surrogate data test linearizes
the series under consideration. Since the histogram of the surrogate
data is identical to the histogram of the original increment series
one can be sure that the probability distribution is not the source of the
nonlinearity of the data. Fig. \ref{fig2} shows the river flow
increment series and its power spectrum before and after the surrogate
data test. Although the river flow increment series exhibits irregular
behavior, its power spectrum shows a very pronounced seasonal peak
with few harmonics. As expected, the surrogate series shows a similar
pattern with almost identical power spectrum.

Next we compare the power spectrum of the volatility series obtained
from the original increment river flow series and from the surrogate
series (Fig. \ref{fig3}). The power spectrum of the original volatility
series shows a pronounced seasonal peak while the power spectrum of
the surrogate volatility series has no seasonal periodicity. The
seasonal periodicity of the original volatility series may be
associated with the increased fluctuation for large river flux (see
Fig. \ref{fig1}). The absence of seasonal periodicity for the
surrogate volatility series is somehow counter intuitive since the
surrogate series itself is as periodic as the original river flow
increment series while a simple inversion operation of the negative
values of $\Delta x_i$ to obtain $|\Delta x_i|$
diminishes this periodicity. The absence of the seasonal periodicity
from the surrogate volatility series indicates that
periodicity in the magnitude series is a result of nonlinearity. We
suggest that the amplitude of the seasonal peak of the original
volatility series compare to the seasonal peak of the surrogate
volatility series would quantify the degree of nonlinearity.

We use the power spectra of the original and surrogate volatility
series to analyze the correlations properties of these series. If a
series $x_i$ is long-range correlated than its autocorrelation
function decays as a power law $C(l) = \frac{1}{N-l}\sum_{i=1}^{N-l}
x_{i+l}x_i \sim l^{-\gamma}$ where $N$ is the series total length, $l$
is the lag, and $\gamma$ is the correlation exponent ($0 < \gamma <
1$). In this case also 
the power spectrum follows scaling law \cite{Shlez,Stanley}
$S(f) \sim 1/f^\beta$ where
$\gamma = 1- \beta$. In Fig.~\ref{fig4} we show the power spectra of
the original and surrogate volatility series for frequencies larger
than $1yr^{-1}$. There is a notable
difference between the power spectrum of the original volatility and
the power spectrum of surrogate
volatility; while the power spectrum of the surrogate volatility
series is almost flat, the power spectrum of the original volatility
decays as a power law with an exponent of $\beta \approx 0.66$. Thus (i) the
original volatility series is power-law correlated and (ii) its
correlations are 
a nonlinear measure since they significantly reduced after the surrogate data
test. The interpretation of these correlations is that there are
clusters of big magnitudes $|\Delta F_i|$ that are statistically
followed by patches of big magnitudes. These clusters are in addition to
the periodic clustering (shown in Fig. \ref{fig3}). We note that
the power spectrum is not the preferred method for scaling analysis;
we repeated the scaling analysis with more advanced
method, the detrended fluctuation analysis \cite{DFA}, and find less
noisy but similar results \cite{remark_DFA}.

We summarize the periodic volatility and the volatility correlations
results for the 30 rivers under consideration in Fig. \ref{fig5}. 
To systematically compare the seasonal periodicity of the different rivers we
first normalize the volatility series by subtracting its mean and dividing
by its standard deviation; thus, the area under the power spectrum of the
different volatility series should be the same. The seasonal peak of
the volatility series exists for all 30 rivers and is significantly
higher then the seasonal peak of the 
surrogate volatility (Fig.~\ref{fig5} upper panel). The scaling
exponent $\beta$ of the original volatility series (Fig. \ref{fig5}
lower panel) indicates correlations; in most of the cases (27/30=90\%)
the exponent of the original volatility series lies above 1 standard
deviation of 
the exponent of the surrogate volatility series. The average $\pm$ 1
standard deviation of the scaling exponent of original volatility
series is $\beta=0.49 \pm 0.11$ and is significantly higher than the
average $\pm$ 1 standard deviation of the scaling exponent of the
surrogate volatility series $\beta=0.18 \pm 0.13$; the $p$ value of
the Student's $t-$test is less than $10^{-6}$. For
time scales larger than 1 year the volatility series is only weakly
correlated with average exponent $\beta=0.27\pm 0.26$.

Thus, we find two measures of nonlinearity related to the river
flow data, periodic volatility and long-range correlated
volatility. These two measures are related to the clustering of the
magnitudes of river flow fluctuations, periodic and long-range
correlated clustering.

To study in more details the possible source for such a seasonal
periodicity of 
the volatility series we propose a simple scheme to generate series
with some similar characteristic properties as in the river flow data. To
mimic the enhanced fluctuations for large river flow we assume that,
\begin{equation} \label{e1}
x_i=(1+A\eta_i)s_i,
\end{equation}
where $\eta_i$ is a Gaussian random variable with zero mean and unit
standard deviation, $A$ is the noise level, and $s_i$ is an
asymmetric periodic function,
\begin{equation} \label{e2}
s_i = s_{j+nT} = \left\{ \begin{array}{lcl}
1+\cos(2\pi fj) &{\rm for}& 0 \le j < \frac{2}{3}T \\
1-\cos(4\pi fj) &{\rm for}& \frac{2}{3}T \le j < T,
\end{array}
\right.
\end{equation}
where $T$ is the time period $T=365$ in arbitrary time units, $j$ is
an integer $0\le j <T$, $f=0.75/T$, and $n$ is an integer. 
$x_i$ decreases for $2/3$ of the time period $T$ and increases for
$1/3$ of this time period.
When the noise level $A$ increases the 
nonlinear term $A\eta_is_i$ also increases. We generate $x_i$ series
with different noise level and then calculate the power spectrum of the
normalized volatility series $|\Delta x_i|$ of the
original and surrogate $\Delta x_i$ series (Fig. \ref{fig7}). We find
that when the noise level is relatively small the seasonal peak is
present in both the original and surrogate volatility series. When the
noise level increases the periodicity of the surrogate
volatility series is diminished. Thus, the larger is the
difference between the peak of original volatility series and
peak of the surrogate volatility series, the larger is the
nonlinearity of $\Delta x_i$. This simple scheme indicates that the
surrogate data test not always deminishes the seasonal periodicity of
the volatility series but rather eliminates the nonlinear part of the
process that is proportional to the noise level (Eq.~\ref{e2}).
We also analyzed time series generated by realistic hydrological model
(ASGi model 
[Kontinuierlicher Abfluss und Stofftransport- Integrierte Modellierung
unter Nutzung von Geoinformationssystemen]
for Bavaria, Germany \cite{ASGi}) for three rivers: Naab, Regniz
and Vils. Both, the seasonal periodicity of the volatility series as well
as its correlations are reproduced by the model and disappear after phase randomization, as 
was observed in the real data.


In summary we analyse the periodic volatilities and the time correlations 
of river flow data for 30 world rivers. We find that
the volatility series are strongly
correlated with a power law behaviour for time scales less than 1 year.
The periodic volatility and the long-range correlated
volatility are found to disappear when randomizing the phases.
This indicate that 
these features of the volatility time series 
are due to nonlinear dynamical processes. We suggest that such
nonlinear features may result from an interaction between noise
and the seasonal trends.

Preliminary analysis of other climate records, such as
daily temperature and daily precipitation records, shows the existence of
periodic and long-range volatility with similar properties as for the
river flow data. Thus, the results presented here may be generic
for other climate records. 



\ifnum\tipo=1
  \figureI
  \figureII
  \figureIII
  \figureIV
  \figureV
  \figureVII
\fi
\ifnum\tipo=2
  \figureI
  \figureII
  \figureIII
  \figureIV
  \figureV
  \figureVII
\fi


\begin{thebibliography}{}
\bibitem{remark_DFA} Unlike the power spectrum, the detrended
fluctuation analysis (DFA) is capable to remove polynomial trends from
the data; see \protect\cite{DFA}. Before applying the DFA method we
filtered out the seasonal periodicity of the data by subtracting the
mean river flow of a specific day in the 365 annual cycle and then
dividing by the standard deviation of this day. Since, in some cases, the
power spectrum of the volatility series of the river flow series contains
harmonics, the scaling exponents obtained by the DFA are more
accurate.
\end{thebibliography}
\end{document}